\documentclass[%
 aip,
 amsmath,amssymb,
 reprint,%
]{revtex4-1}

\usepackage[pdftex]{graphicx}
\usepackage{dcolumn}
\usepackage{bm}

\usepackage[utf8]{inputenc}
\usepackage[T1]{fontenc}
\usepackage{mathptmx}
\usepackage{etoolbox}

\usepackage{amsmath, amssymb, amsthm}
\usepackage{mathtools}
\usepackage{url}
\usepackage{hyperref}
\usepackage{siunitx}

\newcommand{\R}{\mathbb{R}}

\usepackage{color, colortbl}

\makeatletter
\def\@email#1#2{%
 \endgroup
 \patchcmd{\titleblock@produce}
  {\frontmatter@RRAPformat}
  {\frontmatter@RRAPformat{\produce@RRAP{*#1\href{mailto:#2}{#2}}}\frontmatter@RRAPformat}
  {}{}
}%
\makeatother
\begin{document}

\preprint{AIP/123-QED}

\title{Linear Response Theory for Renewable Fluctuations in Power Grids with Transmission Losses}
\author{Anton Plietzsch}
 \altaffiliation[Also at ]{Physics Department, Humboldt-University Berlin}%
 \affiliation{Potsdam-Institute for Climate Impact Research}%
 \email{plietzsch@pik-potsdam.de}%
\author{Sabine Auer}%
 \affiliation{Elena International GmbH}%
\author{Jürgen Kurths}
 \altaffiliation[Also at ]{Physics Department, Humboldt-University Berlin}%
 \affiliation{Potsdam-Institute for Climate Impact Research}%
\author{Frank Hellmann}
 \affiliation{Potsdam-Institute for Climate Impact Research}%

\date{\today}

\begin{abstract}
We study the spreading of renewable power fluctuations through grids with Ohmic losses on the lines. By formulating a network adapted linear response theory, we find that vulnerability patterns are linked to the left Laplacian eigenvectors of the overdamped eigenmodes. We show that for tree-like networks fluctuations are amplified in the opposite direction of the power flow. This novel mechanism explains vulnerability patterns that were observed in previous numerical simulations of renewable micro-grids. While exact mathematical derivations are only possible for tree like networks with homogeneous response, we show that the mechanisms discovered also explain vulnerability patterns in realistic heterogeneous meshed grids by studying the IEEE RTS-1996 test system.
\end{abstract}

\maketitle

\begin{quotation}
Recently, many studies have analysed the spreading of short-term renewable power fluctuations through power grids. In most of these studies, it was assumed that the power transmission on the lines is lossless. For lossless flow networks the flow at the emitting and receiving end of a line are equal. Hence, any flow change will be symmetric on both ends of the line. In contrast, for networks with transmission losses, the flow at the receiving end is always smaller than on the emitting end and changes of the flow at both ends are not symmetric anymore. The spreading of fluctuations through the network will therefore depend on the flow direction at each individual link. Consequently, the nodes that are particularly vulnerable to power fluctuations are not necessarily those that have the strongest excitation for power fluctuations at other nodes. In fact, for renewable fluctuations, we find that all nodes are almost equally excited, while the most vulnerable nodes are located in the high consumption regions in the network, i.e. at the sinks of the power flow.
\end{quotation}

\section{Introduction}

A fundamental challenge for the operation and control of power grids is to maintain the balance between power production and power demand. In AC power systems that are dominated by conventional generators, the power fluctuations on the demand side are balanced by the control schemes of the production side in order to maintain a stable frequency at 50 or 60 Hz, respectively. However, with the ongoing integration of highly intermittent renewable energy sources such as wind and solar there is not only fluctuations on the demand side but also on the generation side \cite{sorensen2007power}. Demand fluctuations are typically uncorrelated and can therefore average out for a large number of consumers. In contrast, the power fluctuations in large wind and solar farms stem from the same meteorological conditions and can therefore be highly correlated. As a result, these fluctuations add up and can lead to large fluctuation of power production at single nodes in the network system \cite{anvari2016short}.


The impact of noise on the stability of the synchronous state in a complex dynamical systems has been intensively studied with the method of linear response theory. Analytical results were given for singular perturbations \cite{tamrakar2018propagation}, white Gaussian noise \cite{Matthiae2016} and exponentially correlated noise \cite{tyloo2018robustness,coletta2018transient, tyloo2019key}. The spreading of intermittency from fluctuations to the frequency response throughout a lossless network was calculated by Haehne et al.\cite{haehne2019propagation}. Zhang et al. identified three frequency regimes of the network response networks: a bulk, a resonant and a local regime. The bulk regime covers low frequency perturbations, and the network responds as a whole. In contrast, as already pointed out by Kettemann et al.\cite{Kettemann2016}, high frequency perturbations stay localized at the fluctuating node and decay exponentially. They constitute the local regime. The resonant regime is where the fluctuation spectrum and the oscillatory dynamics of the network overlap, and produces complex resonant response patterns. 

To our knowledge all prior analytic works on linear response in power systems consider rather simple power system models. In this work we want to transfer this theoretical knowledge and develop a linear response theory that is well suited to also describe more realistic power systems, including higher order dynamical models of inverters or generator. Of particular importance is that our approach is capable of dealing with transport losses on the lines. Mathematically such systems can represented by an asymmetric effective network Laplacian. We derive upper bounds of the response that are highly predictive for the actual behaviour of the system in many cases of interest. The theory is used to explain key features of the complex phenomenology that was numerically observed for renewable fluctuations an AC micro-grid model \cite{auer2017stability}. The major finding is that auto-correlated power fluctuations are enhanced in the opposite direction of the power flow due to Ohmic losses on the lines. With simulations in the IEEE RTS-1996 \cite{grigg1999ieee} test case we are able to show that this mechanism is in fact relevant also for more realistic systems with a meshed topology and heterogeneous parameters. The fact that renewable fluctuations in load heavy regions of the grid have a higher impact on the frequency stability is of high relevance, for example for the connection of new wind parks to a grid.

\section{Power Flow Networks with Losses}

We first give the general form of the power grid equations we want to consider, and derive some general properties of the linearization that help characterize the system.

The network structure of the grid can be represented by a graph $\mathcal{G} = (\mathcal{N,E})$, with a set of $N$ nodes corresponding to generators and loads and a set of $E$ edges corresponding to the transmission lines that carry the power flow. The dynamical state of each node $i$ is represented by $\bm{x}^i(t):\R \to \R^{D_i}$. In the following we will use the notation that node indices are denoted by superscripts and variable indices are denoted by subscripts, such that  $x^i_l$ is the $l$th variable of the $i$th node. The state of the entire dynamical system $\bm{x}(t):\R \to \R^S$ contains the states of all components, with a total system size $S = \sum_{i=1}^N D_i$. We assume that every node $i$ is coupled to its adjacent nodes by a power flow $P^i(\bm{x}):\R^S \to \R$ that depends only the state difference:
\begin{equation}
\label{eq:system}
\begin{split}
    \dot{\bm{x}}^i &= \bm{f}(\bm{x}^i, P^i(\bm{x}))\\
    P^i(\bm{x}) &= \sum_l P^{ij}(x_\theta^i - x_\theta^j)\;.
\end{split}
\end{equation}

Here $P^{ij}(\cdot)$ is the signed power flowing on the line $ij$ as a function of the node states, and as seen from node $i$. If no power is lost on the line we have $P^{ij} = - P^{ji}$. In Eq.~(\ref{eq:system}), we made two additional assumptions:
\begin{enumerate}
    \item The node dynamics is homogeneous, i.e. $\bm{f}^i = \bm{f}^j \eqqcolon \bm{f}$.
    \item The power flow depends only on one internal state variable $x_\theta$, e.g. the voltage phase angle for AC power grids and the absolute voltage for DC power grids.
\end{enumerate}

Further, we require some natural properties to hold for the power flow.
\begin{itemize}
     \item Losses are positive $P_{loss} \coloneqq P^{ij} + P^{ji} > 0$
    \item Positive power flow increases with state difference: If $P^{ij} > 0$ then $\frac{\partial P^{ij}}{\partial x^{ij}_\theta} > 0$.
    \item Losses increase with increasing power flow: If $P^{ij} > 0$ then $\frac{\partial P_{loss}}{\partial x^{ij}_\theta} > 0$.
\end{itemize}

From this it follows that if $P^{ij} > 0$, that is, we have power flowing from $i$ to $j$, we have
\begin{equation}
    \label{eq:weight_relation}
    \frac{\partial P^{ij}}{\partial x^{ij}_\theta} > \frac{\partial P^{ji}}{\partial x^{ji}_\theta} \; .
\end{equation}

We assume the system has a stable stationary state $x = \xi$. The change of the power flow at node $i$ for a deviation $\delta x$ from the stationary state is given by
\begin{equation*}
    \delta P^i = P^i(\xi_\theta + \delta x_\theta) - P^i(\xi_\theta) \approx \sum_j \frac{\partial P^i}{\partial x^j_\theta}(\xi_\theta) \delta x^j_\theta \;.
    \label{eq:power_deviation}
\end{equation*}
We define a matrix $L_{ij} \coloneqq \frac{\partial P^i}{\partial x^j_\theta}(\xi_\theta)$. Inserting the power flow equation we see, that this has the form of a weighted Laplacian matrix
\begin{equation}\label{eq:weighted-laplacian}
    L_{ij} = \delta_{ij} \sum_k w_{ik} - w_{ij},
\end{equation}
with weights $w_{ij} = \frac{\partial P^{ij}}{\partial x^{ij}_\theta}(\xi^{ij}_\theta)$. Usually, Laplacians are defined to be symmetric matrices, with the underlying assumption being a conservation of flow on the links. However, if we consider transport losses, the Laplacian matrix describing the diffusion dynamics on the linear level is asymmetric, i.e. $w_{ij} \neq w_{ji}$.

Similar to the symmetric case, asymmetric Laplacians always have an eigenvalue $\lambda_1 = 0$. The corresponding right eigenvector is homogeneous $\mathrm{v}_{r,i}^{(1)} = \mathrm{v}_{r,j}^{(1)}$ for all $, j$. The corresponding left eigenvector, however, is generally heterogeneous. It is determined by the equation
\begin{equation*}
    0 = \sum_i \mathrm{v}_{l,i}^{(1)} L_{ij} = \sum_i (\mathrm{v}_{l,i}^{(1)} w_{ij} - \mathrm{v}_{l,j}^{(1)} w_{ji}) \eqqcolon \sum_i F_{ij}\;.
\end{equation*}
In tree-like networks this gives a strict relation for the eigenvector entries of two neighbouring nodes. We can see this by starting at the nodes with degree one. At these nodes there is only one summand $F_{ij}$ which therefore has to be zero. All the summands are by definition antisymmetric $F_{ij} = -F_{ji}$ and therefore we know that the corresponding summand $F_{ji}$ in the condition for the neighbouring node is also zero. By going up the tree structure to nodes of higher degree we can eliminate the summands corresponding to all previously visited nodes. Doing this, we see that in the above equations not only the sum is equal to zero but every single summand has to be zero itself and therefore
\begin{equation*}
    \frac{\mathrm{v}_{l,i}^{(1)}}{\mathrm{v}_{l,j}^{(1)}} = \frac{w_{ji}}{w_{ij}}\;.
\end{equation*}
Assume the power is flowing from node $i$ to node $j$. From Eq.~(\ref{eq:weight_relation}) it follows that $w_{ij} > w_{ji}$ and hence, in a tree network the entries of the left eigenvector corresponding to $\lambda_1 = 0$ of the asymmetric Laplacian are increasing along the power flow in the network
\begin{equation}\label{eq:laplacian-ev}
    \mathrm{v}_{l,j}^{(1)} > \mathrm{v}_{l,i}^{(1)} \text{ for } P_{ij} > 0\;.
\end{equation}

Since we assume that the nodes are homogeneous, we can factorize the linearization of the dynamical system (\ref{eq:system}) into a network part and a local part. In Appendix~\ref{sec:jac-factorization} it is shown that the Jacobian of this system can be written in the form
\begin{equation}
\label{eq:Jacobian}
\bm{J} = \bm{A} \otimes \bm{I} + \bm{B} \otimes \bm{L}\;,
\end{equation}
with matrices $\bm{A}, \bm{B} \in \R^{D \times D}$, the weighted Laplacian $\bm{L} \in \R^{N \times N}$ as defined in Eq.~(\ref{eq:weighted-laplacian}) and $\otimes$ denoting the Kronecker product. The eigenvectors and eigenvalues of this Jacobian take the form
\begin{equation}
\label{eq:jac-ev}
\begin{split}
    \bm{v}^{(a,b)} &= \bm{u}^{(a)}(\lambda_a) \otimes \mathrm{v}^{(b)}\;,\\
    \sigma_{(a,b)} &= \mu_a(\lambda_b)\;,
\end{split}
\end{equation}
where $\lambda, \mathrm{v}$ are the eigenvalues and eigenvectors of the Laplacian and $\mu(\lambda), u(\lambda)$ are the eigenvalues and eigenvectors of the matrix $C(\lambda) = A+\lambda B$. In the following, we will use $n$ and $m$ to denote the multi-index $(a,b)$.

\section{Linear Response Theory}


In the following we want to calculate the response of (\ref{eq:system}) to an additive fluctuation $\bm{\eta}(t):\R \to \R^S$. We assume the system has a stable fixed point $\bm{x}(t) = \bm{\xi}$. The linear response of the deviation $\delta \bm{x}(t) \coloneqq \bm{x}(t) - \bm{\xi}$ is then given by
\begin{equation*}
    \delta \bm{x}(t) = \int_{-\infty}^\infty \bm{\chi}(t-t') \bm{\eta}(t') dt'\;,
    \label{eq:response}
\end{equation*}
where $\bm{\chi}(t) = \theta(t)e^{\bm{J} t}$ is the response function of the system. In Fourier space, the convolution reduces to a simple product
\begin{equation*}
    \delta \hat{\bm{x}}(\nu) = \hat{\bm{\chi}}(\nu) \cdot \hat{\bm{\eta}}(\nu) \;.
\end{equation*}
We quantify the response signal by applying the $L_2$ norm. For a single system variable, it is defined as
\begin{equation*}
    \| \delta x_i(t) \|_2 = \sqrt{\int_{-\infty}^\infty |\delta x_i(t)|^2 dt}\;.
\end{equation*}
From Parseval's theorem it follows that $\| \delta x_i(t) \|_2 = \| \delta \hat x_i(\nu) \|_2$. In the following we will restrict ourselves to the case of single node fluctuations, where there is only one non-zero entry $\eta_j(t)$ driving the system. In principle it is straightforward to generalize our approach to fluctuations at multiple nodes. In that case, not only the auto-correlation but also the cross-correlation of fluctuations has to be taken into account. However, the focus on single node fluctuations is sufficient to understand the effects of auto-correlated fluctuations and transport losses. For single node fluctuations the $L_2$ norm of the response is given by
\begin{equation}\label{eq:fourier-response}
    \| \delta x_i(t) \|_2 = \sqrt{\frac{1}{2\pi}\int_{-\infty}^\infty |\hat{\chi}_{ij}(\nu)|^2 S_{\eta_j\eta_j}(\nu) d\nu}\;,
\end{equation}
where $S_{\eta_j\eta_j}(\nu) = |\hat{\eta}_j(\nu)|^2$ is the power spectrum of the fluctuation. The response matrix can be decomposed into the response of the single eigenmodes $\hat{\bm{\chi}}(\nu) = \sum_n \hat{\bm{\chi}}^{(n)}(\nu)$. In Appendix~\ref{sec:mode-decomposition} it is shown that these mode response functions are given by
\begin{equation*}
\hat{\bm{\chi}}^{(n)}(\nu) = \frac{\bm{v}_{r}^{(n)} \bm{v}_{l}^{(n)}}{j\nu-\sigma_n}.
\end{equation*}
where $\sigma_n$ are the eigenvalues of the Jacobian and $v_{r}^{(n)} v_{l}^{(n)}$ is the outer product of the corresponding right and left eigenvectors. Inserting the mode expansion into Eq.~\eqref{eq:fourier-response} yields a sum of single mode terms $|\hat{\chi}_{ij}^{(n)}|^2$ and cross-mode terms $\hat{\chi}_{ij}^{(n)}\bar{\hat{\chi}}_{ij}^{(m)}$. Denoting $\gamma_n = |\Re(\sigma_n)|$ and $\nu_n = \Im(\sigma_n)$, the single mode terms are given by
\begin{equation*}
    |\hat{\chi}_{ij}^{(n)}(\nu)|^2 = \frac{\pi}{\gamma_n} |v_{r,i}^{(n)}|^2|v_{l,j}^{(n)}|^2 L^{(n)}(\nu)\;,
\end{equation*}
where $L^{(n)}$ are Lorentzian functions with width $\gamma_n$ and maximum at $\nu = \nu_n$
\begin{equation*}
    L^{(n)}(\nu) = \frac{1}{\pi}\frac{\gamma_n}{\gamma_n^2 + (\nu-\nu_n)^2}\;.
\end{equation*}
For auto-correlated perturbation signals the integral~\eqref{eq:fourier-response} is generally hard to solve, particularly when the power spectral density is not known analytically. In that case we could only determine the power spectral density from measurements and compute the $L_2$-norm semi-analytically. An analytical approximation for the single mode terms can be calculated for small $\gamma$, the low damping regime. In the limit $\gamma_n \to 0$, the Lorentzian function converge towards a Dirac delta distribution
\begin{equation*}
    \lim_{\gamma_n \to 0} L^{(n)}(\nu) = \delta(\nu-\nu_n)\;,
\end{equation*}
and hence, for small $\gamma_n$ we can approximate the integral as
\begin{equation*}
    \int_{-\infty}^\infty L^{(n)}(\nu) S_{\eta_j\eta_j}(\nu) d\nu \approx S_{\eta_j\eta_j}(\nu_n)\;.
\end{equation*}
This approximation is valid if the spectral density does not vary much over width of the Lorentzian functions $S_{\eta_j\eta_j}(\nu) \approx S_{\eta_j\eta_j}(\nu+\gamma_n)$. In Appendix~\ref{sec:cross-mode-terms} it is further shown that for small damping parameters $\gamma_n$ the cross-mode terms are suppressed. Neglecting these terms is valid if the mode dampings are much smaller than their spectral distance $\gamma_n, \gamma_m \ll |\nu_n-\nu_m|$. The $L_2$ norm of the response can then be approximated as
\begin{equation}\label{eq:peak_approx}
    \| \delta x_i(t) \|_2 \approx \sqrt{\sum_n \frac{1}{2 \gamma_n} |v_{r,i}^{(n)}|^2|v_{l,j}^{(n)}|^2 S_{\eta_j\eta_j}(\nu_n)}\;,
\end{equation}
which we call the \emph{peak approximation}. We see that the response is given by a superposition of the different mode contributions. How strongly a certain mode is excited depends on the power spectral density at the eigenfrequency $\nu_n$ and the entry of the left corresponding eigenvector at the perturbed node, whereas the response strength at different nodes is determined by the entries of the right corresponding eigenvector.

\begin{figure*}[ht!]
	\includegraphics[width=1.0\textwidth]{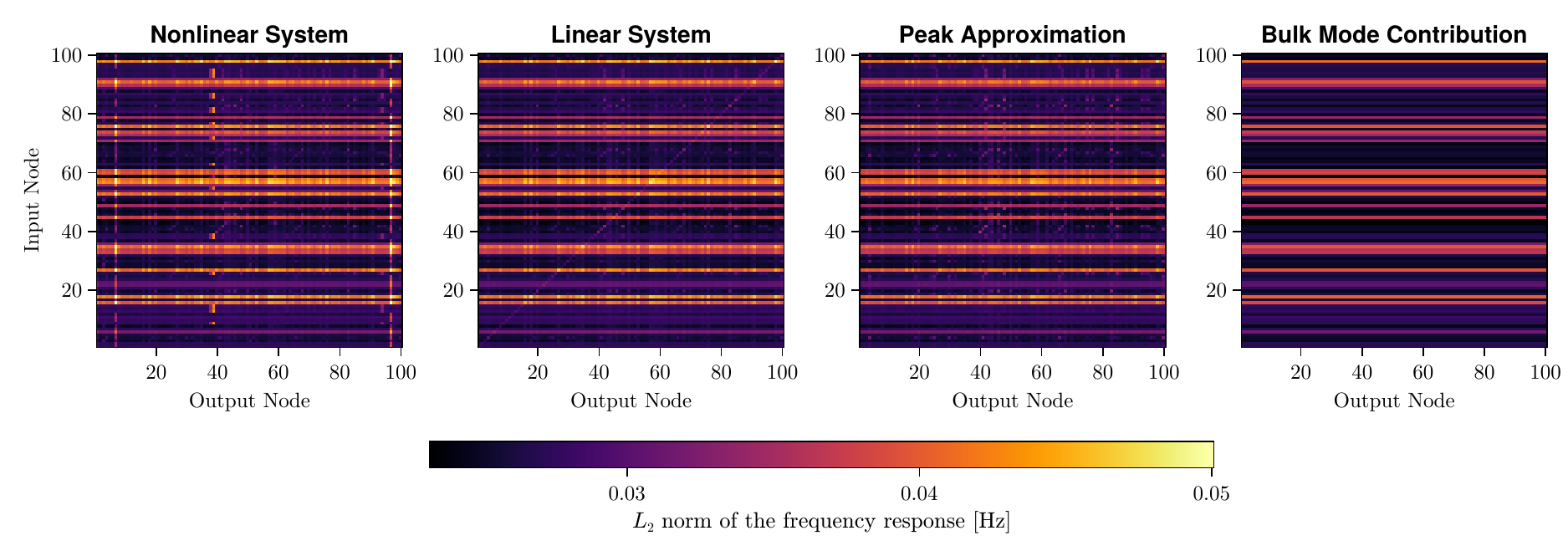}
	\caption{\textbf{Colour plot of the $L_2$ norm of the frequency response at single nodes.} The colour corresponds to the $L_2$ norm of the frequency response at the output nodes (x-axis) given a turbulent fluctuation at the input nodes (y-axis) in the grid depicted in Fig.~\ref{fig:troublemaker_plot}. (a) Simulations of the full nonlinear system for each pair of input and output nodes. (b) Simulation of the linearized system. (c) Analytic prediction calculated with the peak approximation~(\ref{eq:peak_approx}). (d) Contribution of the bulk mode to the analytic prediction.}
	\label{fig:variance_color_plot}
\end{figure*}

\section{Example: AC Micro-Grid Model}

We now will turn towards an example system to demonstrate the the power of our theoretical approach. We analyse the impact of turbulent fluctuations on power grids by characterising the \emph{vulnerability} and \emph{excitability} of the different nodes in the network. Here, we define the vulnerability of a node as the strength of network response to power fluctuations at this particular node. Conversely, the excitability of a node is the strength of the response at this node given a power fluctuations at another node. Using our approximation of the $L_2$ norm we are able to explain three previously observed properties of renewable power grids \cite{auer2017stability}:

\begin{enumerate}
    \item There is a pronounced fine structure in both vulnerability and excitability of nodes.
    \item Losses on the lines lead to a pronounced network structure in the vulnerability of the nodes, but not in which nodes are excited.
    \item The vulnerability appears high in parts of the network that are consumer heavy, and within these areas tends to rise the further away from the center of the network the node is.
\end{enumerate}

In the following we will provide an analytical explanation for these observations for an islanded micro-grid with fluctuating renewable in-feed. Following Schiffer et al.\cite{schiffer2013synchronization}, droop controlled inverters with virtual inertia in their simplest form can be modelled by the swing equation
\begin{equation}
\label{eq:swing_eq}
\begin{split}
\dot \phi_i &= \omega_i\;,\\
M_i \dot \omega_i &= P_i(t) - D_i \omega_i - \sum\limits_{k=1}^N P_{ik}(\phi_i-\phi_k)\;.
\end{split}
\end{equation}
We include resistive losses of the power flow on the lines via the conductance matrix $G_{ik}$
\begin{equation}
    P_{ik} = |V_i||V_k| [G_{ik} \cos(\phi_i-\phi_k) + B_{ik} \sin(\phi_i-\phi_k)]\;.
\end{equation}
Further, we assume that the power in-feed at each node is composed of a constant and a small fluctuating part
\begin{equation}
    P_i(t) = P_i + \delta P_i(t)\;.
\end{equation}
We simulate this system on a 100 node network generated by a random growth model for power grids \cite{Schultz2014}. The power fluctuation signal is generated by a combination of stochastic wind and solar power fluctuation models \cite{Schmietendorf2017, Anvari2017}. The power spectrum of the resulting signal is power-lawed with the Kolmogorov exponent of turbulence. Further details on the parametrization and the fluctuation modelling can be found in Appendix~\ref{sec:micro-grid-parametrization}. The response at each node is quantified in terms of the $L_2$ norm of the frequency deviation from the nominal grid frequency $\omega_s$
\begin{equation*}
    \| \delta \omega_i(t) \|_2 = \sqrt{\int_{-\infty}^\infty (\omega_i(t)-\omega_s)^2 dt}\;.
\end{equation*}
The response of the entire dynamical system $\mathcal{S}$ can then be quantified in terms of the $L_2$ norm of the average deviation from the nominal grid frequency
\begin{equation}\label{eq:dev-norm}
    \|\mathcal{S}\|_\text{dev} = \sqrt{\int_{-\infty}^\infty \frac{1}{N}\sum_{i=1}^N(\omega_i(t) - \omega_s)^2 dt}\;.
\end{equation}
It should be noted, that this measure is different from the synchronization norm that has been used in most of the studies on the response of swing equations in the linear regime~\cite{andreasson2017coherence,tyloo2018robustness,tyloo2019key,totz2020control}
\begin{equation*}
        \|\mathcal{S}\|^2_\text{sync} = \sqrt{\int_{-\infty}^\infty\frac{1}{N}\sum_{i=1}^N\left(\omega_i(t) - \frac{1}{N}\sum_{l=1}^N\omega_l\right)^2 dt}\;.
\end{equation*}
While being useful to study the synchonicity in the network, this measure by definition omits any fluctuation of the bulk of synchronous frequencies. However, as will be shown in the following, this bulk behavior turns out to be the most dominant mode in the frequency response to renewable power fluctuations. Further, due to the presence of losses, this mode is no longer homogeneous throughout the network. As we will see it can completely dominate the effect of network structure on the systems node wise vulnerability.

\subsection{Fine Structure of Network Responses}


Following Auer et al. \cite{auer2017stability}, we simulate single node fluctuations in the full nonlinear system~(\ref{eq:swing_eq}) for every pair of perturbed (input) and observed (output) nodes and depict the response strength as a color coded matrix plot (Fig.~\ref{fig:variance_color_plot}). There, the horizontal and vertical lines correspond to nodes with large vulnerability or excitability, respectively. Comparing the simulation of the linearized system with the full nonlinear system shows that the main response pattern remains also in the linearized dynamics. However, there are some nonlinear artefacts that are not captured by the linearized model. These nonlinear effects can obviously not be analysed with our linear theory. For a given time series of a power fluctuation $\delta P(t)$, we can numerically determine its power spectral density and thereby semi-analytically compute the peak approximation~(\ref{eq:peak_approx}) for the $L_2$ norm of the frequency responses. In Fig.~\ref{fig:variance_color_plot} it can clearly be seen that this approximation is able to reproduce the response pattern of the linearized system. This means that by only knowing the systems eigenvectors and the spectral density of the power fluctuations at the systems eigenfrequencies, we are able to analytically predict the response strength at every node in the network. The full network response is a superposition of the responses of every single mode. The contribution of each eigenmode is determined by the spectral excitation factor and the response pattern by the left and rights eigenvectors of that mode. When a single mode is dominating the response of the network, the vulnerability and excitability of the nodes can be linked to the left and right eigenvectors of this mode.

\subsection{Line Losses and the Bulk Mode}

In the following we will assume homogeneous damping and inertia parameters $D_i = D, \ M_i = M$. The Jacobian of the system~(\ref{eq:swing_eq}) can then be written in the form (\ref{eq:Jacobian}), with matrices
\begin{equation*}
\bm{A} =
\begin{pmatrix}
0 & 1 \\
0 & -\frac{D}{M}
\end{pmatrix}, \quad
\bm{B} = 
\begin{pmatrix}
0 & 0 \\
-\frac{1}{M} & 0
\end{pmatrix},
\end{equation*}
and the Laplacian weights
\begin{equation*}
    w_{ik} = |V_i||V_k| [B_{ik} \cos(\phi_{ik}) - G_{ik} \sin(\phi_{ik})]\;.
\end{equation*}
From Eq.~(\ref{eq:jac-ev}) it follows that
\begin{equation}
    \label{eq:swing_eigvals}
    \sigma_{(\pm,b)} = -\frac{D}{2M} \pm \sqrt{\frac{D^2}{4M^2}-\frac{\lambda_b}{M}}\;.
\end{equation}
For the Laplacian eigenvalue $\lambda_1 = 0$ we have two Jacobian eigenvalues, $\sigma_{(+,1)}=0$ and $\sigma_{(-,1)}=-\frac{D}{M}$. The eigenvalue $\sigma_{(+,1)}$ corresponds to the symmetry of homogeneous phase shifts that do not contribute to the dynamics, whereas $\sigma_{(+,1)}$ corresponds to homogeneous frequency shifts leading to an exponentially decaying response of the nodes' frequencies with rate $\frac{D}{M}$. When the algebraic connectivity of the network, i.e. the second smallest eigenvalue of the Laplacian, fulfils the condition $\lambda_2 > \frac{D^2}{4M}$, the square root term in (\ref{eq:swing_eigvals}) will be imaginary and therefore, $\sigma_{(-,1)}$ is the only overdamped mode in the system. In this case the mode fully determines the behaviour of the system in the bulk regime and we therefore refer to it as the bulk mode.

When the algebraic connectivity is significantly larger than the threshold $\lambda_2 \gg \frac{D^2}{4M}$, the eigenfrequencies of all the other system modes are rather high. For correlated fluctuations the power spectrum at high frequencies is suppressed~\cite{milan2013turbulent,anvari2016short} and therefore we find that the network response in this regime is entirely dominated by the bulk mode. The right Laplacian eigenvector of this mode is homogeneous whereas the left eigenvector has heterogeneous entries. This means that all nodes are equally excited but certain nodes have much higher vulnerability to power fluctuations. The resulting dynamical asymmetry corresponds to the continuous horizontal lines in the bulk mode plot (Fig.~\ref{fig:variance_color_plot}) and has its origin in the Ohmic losses of the lines.

\begin{figure}[htb]
	\includegraphics[ width=\columnwidth]{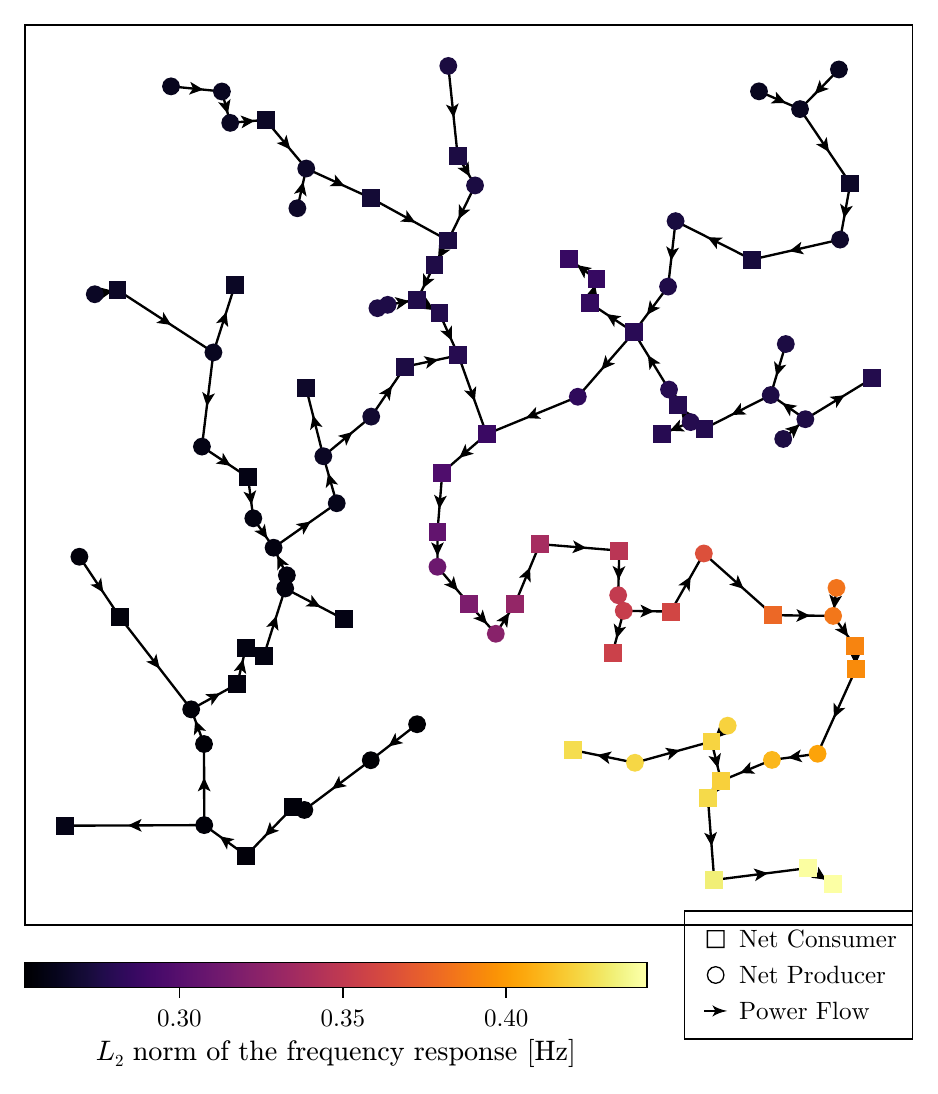}
	\caption{\textbf{Vulnerability of nodes in the power flow network for bulk dominated dynamics.} Nodes are coloured according to the deviation norm (\ref{eq:dev-norm}) of the system response for fluctuations at this node. The link arrows indicate the direction of the power flow.}
\label{fig:troublemaker_plot}
\end{figure}

\subsection{Vulnerability in Tree-Networks}

When the bulk mode is dominating the response of the system, the difference of the vulnerability of nodes to power fluctuations is entirely determined by the left eigenvector of this mode. From Eq.~(\ref{eq:laplacian-ev}) it follows, that entries of this eigenvector are increasing along the power flow. This means that fluctuations are amplified in the opposite direction of the power flow and the nodes located at the sinks of the flow are the most vulnerable. In Fig.~\ref{fig:troublemaker_plot} it can clearly be seen that the network branch where the power is flowing from the center towards the outlying nodes is much more vulnerable than the network branches where the power is flowing towards the center. This explains the observation by Auer et al. \cite{auer2017stability} that the vulnerability of nodes and the closeness centrality of the network are closely related.

\section{Example: IEEE Reliability Test System-1996}

As a last example, we simulate the IEEE Reliability Test System-1996 \cite{grigg1999ieee} to show that our findings are still valid in a much more realistic test case, including a meshed grid topology, (algebraicly modelled) load buses and heterogeneous generator parameters. Similar to the example in the previous section we simulate single node power fluctuations using a combination of stochastic wind and solar power fluctuation models \cite{Schmietendorf2017, Anvari2017}. Imagine a large solar or wind farm connected to one of the buses in the system. Since the fluctuations at single solar panels and wind turbines can be highly correlated within a farm, these fluctuations will not average out but rather add up to a large power fluctuation at the respective node.

\begin{figure}[ht!]
	\includegraphics[width=\columnwidth]{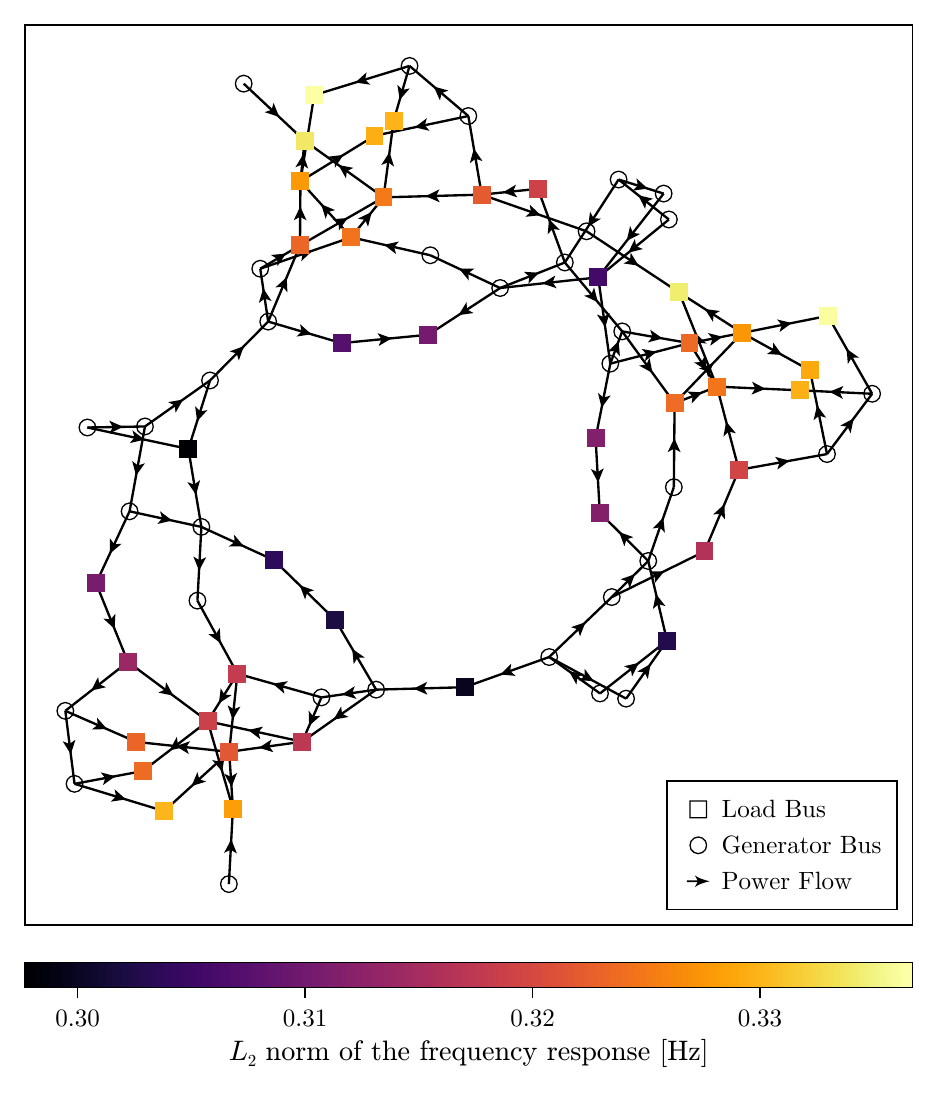}
	\caption{\textbf{Vulnerability of load nodes to intermittent fluctuations in the IEEE Reliability Test System-1996.} Load nodes are depicted as circles and coloured according to the $L_2$-norm of the system response for fluctuations at this node. Generator nodes are depicted as squares. The link arrows indicate the direction of the power flow.}
	\label{fig:ieee96}
\end{figure}

We simulate single node power fluctuations at the load buses and measure the frequency response at the generator buses in the system using the deviation norm~(\ref{eq:dev-norm}). The results are depicted in Fig.~\ref{fig:ieee96}. It can clearly be seen that the vulnerability towards power fluctuations at a bus increases in the direction of the power flow. Accordingly, the must vulnerable buses are located in the parts of the grid with a very high share of loads. This indicates that our findings concerning the joined effect of auto-correlated power fluctuations and line losses are also valid in a more realistic system setup.

\section{Conclusion}

In this paper we presented a linear response theory for correlated fluctuations in power grids that was used to derive an approximation for the frequency response that turned out to be highly accurate in the regime studied. We have shown that important features of the response can be understood as a consequence of the Laplacian that describes the dynamical coupling of nodes in the network being asymmetric in the presence of Ohmic losses on the power lines. In particular, we were able to fully explain the structures in the node vulnerability that have been observed in numerical simulations of an islanded micro-grid with high renewable penetration \cite{auer2017stability}. For tree-like grids the location of vulnerable nodes is related to the power flow throughout the network. In particular, fluctuations at net power flow sinks result in a strong frequency response at all nodes in the network. For tree-like grids with a very unbalanced power production, we find that the consumer heavy branches are much more vulnerable to turbulent in-feed of renewable power. This effect is direct consequence of both the losses on the power lines and the correlated nature of renewable power fluctuations. Considering the generality of our theoretical approach, it should be mentioned that the application to power grids is not limited to fluctuation of renewable energy sources but might also be the basis for studying the impact of demand fluctuations on grids. Also, this work was focused on single node fluctuations. However, the formulation of the response in terms of power spectra also provides an elegant starting point for understanding correlated multi-node fluctuations. In that case not only the auto-correlation but also the cross-correlation of fluctuations will play a crucial role for understanding the dynamical interactions. Finally, the formulation used here is suited to study active power flow variations. To understand the response of the full active and reactive power flow, as well as voltage variations, e.g. for the study of fluctuations in effective models \cite{kogler2022normal}, the approach needs to be generalized to more general models than Eq.~\eqref{eq:system}.

\begin{acknowledgments}
The work presented was partially funded by the BmBF (Grant No. 03EK3055A) and the Deutsche Forschungsgemeinschaft (Grant No. KU 837/39-1 / RA 516/13-1).
\end{acknowledgments}

\section*{Author Declarations}

\subsection*{Conflict of Interest}

The authors have no conflicts to disclose.

\subsection*{Author Contributions}

\textbf{Anton Plietzsch:} Conceptualization (equal); Data curation (equal); Formal analysis (lead); Investigation (lead); Methodology (equal); Software (equal); Visualization; Writing – original draft (lead); Writing – review \& editing (equal).
\textbf{Sabine Auer:} Conceptualization (equal); Data curation (equal); Methodology (equal); Software (equal); Writing – original draft (supporting).
\textbf{Jürgen Kurths:} Conceptualization (equal); Funding acquisition (equal); Supervision; Writing – review \& editing (equal).
\textbf{Frank Hellmann:} Conceptualization (equal); Formal analysis (supporting); Funding acquisition (equal); Investigation (supporting); Methodology (equal); Project administration; Writing – original draft (supporting); Writing – review \& editing (equal).

\section*{Data Availability Statement}

Data and code for reproducing the figures is available at the GitHub repository~\url{https://github.com/PIK-ICoNe/LinearResponsePaper}.

\appendix

\section*{Appendix}

\section{Facorization of the Network Jacobian}\label{sec:jac-factorization}

Taking the total derivative of the $k$th component of the right-hand-side function at the $i$th node with respect to the $l$th variable at the $j$th node yields
\begin{equation*}
    \frac{df_k^i}{dx_l^j} = \frac{\partial f_k^i}{x_l^i}\delta_{ij} + \frac{\partial f_k^i}{\partial p^i}\delta_{l\theta}  \left(\delta_{ij}\sum_n \frac{\partial p^{in}}{\partial x_\theta^i} - \frac{\partial p^{ij}}{\partial x_\theta^i}\right)\;.
\end{equation*}
Defining the matrices
$A_{kl} = \frac{\partial f_k}{x_l}\;, \quad B_{kl} = \frac{\partial f_k}{\partial p} \delta_{l\theta}$ and using the fact that the definition of the weighted Laplacian~\eqref{eq:weighted-laplacian} yields the Jacobian~(\ref{eq:Jacobian}). For a right eigenvector of this Jacobian we have:
\begin{equation*}
\begin{split}
\bm{J} v^{(a,b)} &= (\bm{A} \otimes \bm{I} + \bm{B} \otimes \bm{L}) \bm{u}^{(a)}(\lambda_b) \otimes \mathrm{v}^{(b)}\\
&= \bm{A} \bm{u}^{(a)}(\lambda_b) \otimes \mathrm{v}^{(b)} + \bm{B} \bm{u}^{(a)}(\lambda_b) \otimes \bm{L} \mathrm{v}^{(b)}\\
&= \bm{A} \bm{u}^{(a)}(\lambda_b) \otimes \mathrm{v}^{(b)} + \lambda_b \bm{B} \bm{u}^{(a)}(\lambda_b) \otimes \mathrm{v}^{(b)}\\
&= (\bm{A} + \lambda_b \bm{B}) \bm{u}^{n}(\lambda_b) \otimes \mathrm{v}^{(b)}\\
&= \mu_a(\lambda_b) \bm{v}^{(a,b)} = \sigma_{(a,b)} \bm{v}^{(a,b)}
\end{split}
\end{equation*}
The proof for the left Jacobian eigenvectors can be done similarly.

\section{Mode Decomposition of the Response Function}\label{sec:mode-decomposition}

The Fourier transform of the response function $\bm{\chi}(t) = \theta(t) e^{J t}$ is given by
\begin{equation*}
    \hat{\bm{\chi}}(\nu) = (j\nu \bm{I} - \bm{J})^{-1}\;.
\end{equation*}
The Jacobian can be factorized as $\bm{J} = \bm{Q} \Sigma \bm{Q}^{-1}$, where $\bm{Q}$ and $\bm{Q}^{-1}$ are given by the left and right eigenvectors
\begin{equation*}
\bm{Q} =
\begin{bmatrix}
\bm{v}_r^{(1)} & \dots & \bm{v}_r^{(n)}
\end{bmatrix}\;,
\quad
\bm{Q}^{-1} =
\begin{bmatrix}
\bm{v}_l^{(1)} \\ \vdots \\ \bm{v}_l^{(n)}
\end{bmatrix}\;,
\end{equation*}
and $\Sigma$ is a diagonal matrix whose diagonal elements are the eigenvalues $\Sigma_{nn} = \sigma_n$. Using this, we can show that
\begin{equation*}
\begin{split}
(j\nu \bm{I} - \bm{J})^{-1} &= \bm{Q} \bm{Q}^{-1} (j\nu \bm{I} - \bm{J})^{-1} \bm{Q} \bm{Q}^{-1} \\
&= \bm{Q} (j \nu \bm{I} - \bm{Q}^{-1} \bm{J} \bm{Q})^{-1} \bm{Q}^{-1} \\
&= \bm{Q} (j \nu \bm{I} - \bm{\Sigma})^{-1} \bm{Q}^{-1} \;.
\end{split}
\end{equation*}
An element of this matrix is given by
\begin{equation*}
[\bm{Q} (j \nu I - \Sigma)^{-1} \bm{Q}^{-1}]_{ij} = \sum_{n=1}^N \frac{Q_{in}Q^{-1}_{nj}}{j \nu - \sigma_n} = \sum_{n=1}^N \frac{\bm{v}_{r,i}^{(n)}\bm{v}_{l,j}^{(n)}}{j \nu - \sigma_n} \;,
\end{equation*}
and therefore
\begin{equation*}
\hat{\bm{\chi}}(\nu) = \sum_{n=1}^N \frac{\bm{v}_r^{(n)}\bm{v}_l^{(n)}}{j \nu - \sigma_n} \;.
\end{equation*}

\section{Suppression of Cross-Mode Terms}\label{sec:cross-mode-terms}

The the contribution of cross-mode terms is quantified by
\begin{equation*}
    \int_{-\infty}^\infty \hat{\chi}_{ij}^{(n)}\bar{\hat{\chi}}_{ij}^{(m)} d\nu = \int_{-\infty}^\infty \frac{v_{r,i}^{(n)}v_{l,j}^{(n)} \bar{v}_{r,i}^{(m)}\bar{v}_{l,j}^{(m)}}{(\nu-\nu_n + j\gamma_n)(\nu-\nu_m - j\gamma_m)}d\nu\;.
\end{equation*}
We define the function
\begin{equation*}
    f^{(n,m)}(\nu) = \frac{1}{(\nu-\nu_n + j\gamma_n)(\nu-\nu_m - j\gamma_m)}\;,
\end{equation*}
The integral of this function can be solved using the residue theorem
\begin{equation*}
    \begin{aligned}
        \int_{-\infty}^\infty f^{(n,m)}(\nu) d\nu
        &= 2\pi j \ \text{Res}(f^{(n,m)},\nu_m+j\gamma_m)\\
        &= \frac{2\pi j}{\nu_m-\nu_n + 2j \gamma_{nm}}\;,
    \end{aligned}
\end{equation*}
where we defined $\gamma_{nm} = (\gamma_n+\gamma_m)/2$. For single mode terms ($n=m$), the integral is growing by a factor $\gamma_n^{-1}$ as $\gamma_n \to 0$, whereas for the cross-mode terms ($n \neq m$) the integral converges to a finite constant. Hence, if the damping parameters $\gamma_{nm}$ are significantly small, the cross-mode terms will be suppressed. Moreover, the cross-mode term $\hat{\chi}_{ij}^{(n)}\bar{\hat{\chi}}_{ij}^{(m)}$ will be even smaller, if the overlap of the eigenvectors of modes $n$ and $m$ is small. For systems where factorize into network and internal parts (see Eq.~\eqref{eq:jac-ev}), this is the case when the Laplacian eigenvectors have little overlap in their support on the network. While it is hard to give general rules when this will be the case, this could explain why the approximation in Eq.~\eqref{eq:peak_approx} gives reasonable results even outside of the low-damping limit.

\section{Parametrization of the Micro-Grid Model}\label{sec:micro-grid-parametrization}

The micro-grid model case is kept at a conceptual level to study the effect of local fluctuations on dynamic grid stability and isolate the influence of the network structure. Germany has 4,500 MV distribution networks that connect 500,000 LV distribution networks \cite{bossmann2015shape}. Thus the micro-grid is chosen as a network of 100 nodes to represent an average German grid at medium-voltage (MV) level. The MV level is a good testing case for modelling power grids with a high renewable energy share, since most PV power plants are connected to low-voltage- (LV) or MV levels.  An islanded micro-grid must be internally power-balanced and not connected to a higher grid level. Being power balanced, we assume that there are 50 net producers and 50 net consumers with $P_i=\pm \SI{0.2}{MW}$ power in-feed before losses. The power in feeds are chosen homogeneously to focus on topology and network effects in the model. As there is no connection to upper grid levels, losses are compensated locally at each node, and the net power in-feed is given by $\tilde{P}_i = (P_i +P_{loss}/N)$. Mathematically this is equivalent to switching to the co-rotating frame. The network topology is generated by a random growth model for power grids \cite{Schultz2014}. We have chosen a parametrization such that we get tree-like grids which is a typical structure for distribution grids. The grid parametrization follows from the voltage level. The line impedance for typical MV grids with \SI{20}{kV} base voltage equals $Z = R+jX \approx \SI[parse-numbers=false]{(0.4 + 0.3j)}{\Omega/km}$ \cite{auer2016can}. For simplicity all power, voltage and impedance values are transformed into per unit with a base voltage of \SI{20}{kV} and a base power of \SI{1}{MW}, which are typical values for MV grids \cite{auer2016can,sen2007principles}. The absolute impedance of each line scales with the geographic distance $l$ between linked nodes and consequently differs per link. The average line length is \SI{23.7}{km}\cite{auer2016can}. Further, the model case is assumed to be dominated by inverters to analyze a scenario with high renewable penetration. Wind and solar power plants are connected to the grid via inverters. In an islanded scenario some of these inverters will need to be grid forming to ensure frequency stability. As mentioned above, network nodes are aggregated with a mix of grid-forming inverters, grid-feeding inverters and demand \cite{schiffer2016survey}. Since grid-feeding inverters do not contribute any inertia, the effective nodes have inertia much lower than nodes fully consisting of grid-forming inverters would have. Grid-forming inverters are modeled following Schiffer et al. \cite{schiffer2013synchronization} with a droop controlled frequency based on a low pass filtered power measurement. The virtual inertia and damping for the network model is then given by the low-pass filter exponent $\tau_p$ and the droop control parameter $k_p$ from grid-forming inverters: $M = \tau_p/k_p$, $D=1/k_p$, $\forall i$ with $i=1,\dots,N$. Standard parameters for the droop and time constants of grid-forming inverters are in the range $k_p= \SIrange[range-phrase = \dots]{0.1}{10}{s^{-1}}$ and $\tau_p= \SIrange[range-phrase = \dots]{0.1}{10}{s}$ \cite{schiffer2013synchronization,coelho2002small}. For the simulations we therefore used $M = \SI{0.1}{s}$ and $D = \SI{0.01}{s^2}$. In the simulations intermittent time series for solar and wind power fluctuations were generated by a clear sky index model, based on a combination of a Langevin and a jump process, developed by Anvari et al. \cite{anvari2016short}, and a Non-Markovian Langevin type model developed by Schmietendorf et al. \cite{Schmietendorf2017}, respectively. The power fluctuation $\delta P(t)$ is a combination of the signals generated with these models for wind and solar power fluctuations
\begin{equation}
\Delta d(t) = 0.5 \delta P_W(t)+0.5 \delta P_S(t).
\end{equation}
Both the PDF of the fluctuations and their increment time series are fat tailed (the tails are not exponentially bounded \cite{asmussen2008applied}). The power generation from wind and solar power plants has a power spectrum that is power-lawed with the Kolmogorov exponent of turbulence \cite{milan2013turbulent,anvari2016short}. Thus, these time series show long-term temporal correlations.

\bibliographystyle{apsrev4-1}
\bibliography{bibliography}

\end{document}